\documentclass[fleqn,10pt]{article} 
\usepackage{graphicx,url}
\usepackage[utf8x]{inputenc}
\usepackage{longtable,color}
\usepackage[top=3cm, bottom=3cm, left=3cm, right=2cm]{geometry}
\usepackage{verbatim}
\usepackage{color}
\usepackage{xspace}
\usepackage{rotating}
\usepackage{times}
\usepackage{latexsym}
\usepackage{amstext}
\usepackage{enumerate}
\definecolor{codegreen}{rgb}{0,0.4,0}
\definecolor{codeblue}{rgb}{0,0.5,1}
\definecolor{codegray}{rgb}{0.5,0.5,0.5}
\definecolor{codepurple}{rgb}{0.58,0,0.82}
\definecolor{backcolour}{rgb}{0.95,0.95,0.92}

\usepackage[noend, vlined, linesnumbered, ruled, english]{algorithm2e}

\usepackage[scaled=0.9]{DejaVuSansMono}
\usepackage[T1]{fontenc}

\usepackage{listings}
\lstdefinestyle{mystyle}{
    commentstyle=\color{codegreen},
    keywordstyle=\color{magenta},
    numberstyle=\tiny\color{codegray},
    stringstyle=\color{codepurple},
    basicstyle=\small\ttfamily,
    breakatwhitespace=false,         
    breaklines=true,                 
    captionpos=b,                    
    keepspaces=true,                 
    showspaces=false,                
    showstringspaces=false,
    showtabs=false,                 
    tabsize=2
}
\lstdefinelanguage{SwiftK}
{
	alsoletter={<, :, >, /},
	morekeywords={app, stdout, foreach, in, filename},
    basicstyle=\footnotesize\ttfamily,
    keywordstyle=\color{codeblue},
    numberstyle=\tiny\color{codegray},
    stringstyle=\color{codepurple},
	morestring=[b]"
}

\title{BioWorkbench: A High-Performance Framework for Managing and Analyzing Bioinformatics Experiments}

\author{Maria~Luiza~Mondelli$^1$ \and Thiago~Magalhães$^1$ \and Guilherme~Loss$^1$ \and Michael~Wilde$^2$ \and Ian~Foster$^2$ \and Marta~Mattoso$^3$ \and Daniel~S.~Katz$^4$ \and Helio~J.~C.~Barbosa$^{1,5}$ \and Ana~Tereza~R.~Vasconcelos$^1$ \and Kary~Ocaña$^1$ \and Luiz~M.~R.~Gadelha~Jr.$^{1,\otimes}$
}
\date{\small
$\mbox{}^1$National Laboratory for Scientific Computing, Petrópolis, Brazil\\
$\mbox{}^2$Computation Institute, Argonne National Laboratory/University of Chicago, Chicago, USA\\
$\mbox{}^3$Computer and Systems Engineering Program, Federal University of Rio de Janeiro, Brazil\\
$\mbox{}^4$National Center for Supercomputing Applications, University of Illinois, Urbana, USA\\
$\mbox{}^5$Federal University of Juiz de Fora, Minas Gerais, Brazil\\
$\mbox{}^\otimes$Corresponding author: {\tt lgadelha@lncc.br}
}

\begin{document}

\flushbottom
\maketitle

\begin{abstract}
Advances in sequencing techniques have led to exponential growth in biological data, demanding the development of large-scale bioinformatics experiments. 
Because these experiments are computation- and data-intensive, they require high-performance computing (HPC) techniques and can benefit from specialized technologies such as Scientific Workflow Management Systems (SWfMS) and databases. 
In this work, we present BioWorkbench, a framework for managing and analyzing bioinformatics experiments. This framework automatically collects provenance data, including both performance data from workflow execution and data from the scientific domain of the workflow application. Provenance data can be analyzed through a web application that abstracts a set of queries to the provenance database, simplifying access to provenance information.
We evaluate BioWorkbench using three case studies: SwiftPhylo, a phylogenetic tree assembly workflow; SwiftGECKO, a comparative genomics workflow; and RASflow, a RASopathy analysis workflow. We analyze each workflow from both computational and scientific domain perspectives, by using queries to a provenance and annotation database. Some of these queries are available as a pre-built feature of the BioWorkbench web application. Through the provenance data, we show that the framework is scalable and achieves high-performance, reducing up to 98\% of the case studies execution time.
We also show how the application of machine learning techniques can enrich the analysis process. 
\end{abstract}

\section{Introduction}\label{sec:intro}


Genome sequencing methodologies have led to a significant increase in the amount of data to be processed and analyzed by Bioinformatics experiments. 
Consequently, this led to an increase in the demand for their scalable execution.
Most bioinformatics studies aim to extract information from DNA sequences and can be classified as \textit{in silico} experiments.
\textit{In silico} comprise mathematical and computational models that simulate real-world situations; 
they depend on the use of computational resources and specialized technologies for their execution. 
Simulations often require the composition of several applications, or activities, which have dependencies and manipulate large amounts of data.
These aspects make it difficult to manage and control \textit{in silico} experiments.
Due to the complexity of simulation-based scientific experiments, it is necessary to use approaches that support their design, execution, and analysis, such as scientific workflows \cite{Deelman2009}. 
A scientific workflow is an abstraction that formalizes the composition of several activities through data set production and consumption. 
Each activity corresponds to a computational application, and the dependencies between them represent the execution data flow, in which the output of one activity is input to another.

Scientific Workflow Management Systems (SWfMS) can be used to manage the various steps of the life-cycle of a scientific workflow, i.e. design, execution, and analysis. 
SWfMS can be deployed in high-performance computing (HPC) environments, such as clusters, clouds, or computational grids.
They can also capture and store provenance information.
Provenance records the process of deriving data from the execution of a workflow. For this reason, provenance describes the history of a given experiment, ensuring its reliability, reproducibility, and reuse \cite{Freire2008}.
Provenance may also contain computational and scientific domain data, making it an important resource for the analysis of the computational behavior of an experiment and its scientific results.
By combining computational and domain data, it is possible, for example, to make optimizations in the experiment as well as predictions for execution time and the amount of storage that it will require.

In this work, we present BioWorkbench: a framework that couples scientific workflow management and provenance data analytics for managing bioinformatics experiments in HPC environments. 
BioWorkbench integrates a set of tools that cover the process of modeling a bioinformatics experiment, including a provenance data analytics interface, transparently to the user. 
For managing scientific workflows, we use SWfMS Swift \cite{Wilde2011}, because of the support that the system provides for executing workflows in different HPC environments transparently and also due to its scalability \cite{Wilde2009}. 
Provenance data related to workflow performance and resulting data associated with the application area of the experiment are automatically collected by the framework.
As part of the framework, we have developed a web application where workflow results are presented in a data analytics interface, from the abstraction of a set of queries to the provenance database, supporting the process of analysis by the scientists. 
We also demonstrate that it is possible to use machine learning techniques to extract relations from provenance data that are not readily detected, as well as to predict and classify the execution time of workflows.
In this way, BioWorkbench consolidates various activities related to scientific workflow execution analysis in a single tool.

We evaluated the BioWorkbench using three case studies: SwiftPhylo, a phylogenetic tree assembly workflow; SwiftGECKO, a comparative genomics workflow; and RASflow, a RASopathy analysis workflow. 
Each case study has its own characteristics. However, the three allowed the evaluation of aspects related to performance gains and provenance management covered by our framework. 
We show results where we obtained a reduction of up to 98.9\% in the total execution time, in the case of SwiftPhylo, decreasing from $\sim$13.35 hours to $\sim$8 minutes. Also, we demonstrate that with the provenance collected through the framework, we can provide useful results to the user who does not need to inspect files manually to extract this information. 

\section{Related Work}\label{sec:related}

In general, our proposal addresses aspects related to the modeling of bioinformatics workflows, their parallel execution in high-performance computing environments, as well as provenance data analytics, including predictions on computational resource usage. Here we compare BioWorkbench to related solutions from these different points of view.


There are a variety of SWfMS for modeling and executing workflows in different application areas. Some of them allow execution in high-performance computing environments. Among them, we highlight Pegasus~\cite{Deelman2015}, which enables the specification of workflows through the XML format. The provenance is managed through the Wings/Pegasus framework \cite{Wings2008}, and can be queried using the SPARQL language.
Askalon \cite{Nadeem2007} allows the definition of the workflow through a graphical interface, using UML diagrams or the XML format. Workflow performance monitoring information is used by a service that supports planning the execution of activities. 
Taverna \cite{Wolstencroft2013} is an SWfMS widely used by the bioinformatics community, where workflow activities usually comprise web services. Parallel execution has an optimization engine that identifies and simplifies complex parts of the workflow structure \cite{Cohen2014}.
Provenance is collected and stored in a database and workflows can be shared through the \textit{myExperiment}\footnote {https://www.myexperiment.org/home} platform. 
With a focus on cloud environments, SciCumulus \cite{silva2014} is a SWfMS based on relational algebra for workflow definition that uses a provenance database for configuration and monitoring of executions. It is distinct in that it allows provenance queries to be made during the execution of the experiment. Compared with these solutions, we use the Swift SWfMS because it transparently supports the execution of workflows in different HPC environments, it has been shown that it has a high potential for scalability \cite{Wilde2009}, and it supports provenance management \cite{Gadelha2012}. In addition, it supports workflow patterns such as conditional execution and iterations that not supported by similar systems such as Pegasus \cite{Deelman2015}. It also evaluates the workflow dynamically, possibly changing task execution sites for efficiency of reliability.

Juve et al. \cite{Juve2013} present Wfprof, a tool for collecting and summarizing performance metrics of workflow activities. Wfprof was coupled to the Pegasus SWfMS to identify the complexity level of different workflows and how computational and I/O intensive they are. 
In \cite{Krol2016}, an approach for analyzing the performance of workflows executed with the Pegasus SWfMS is presented, which also studies the effect of input parameters on task performance. 
The PDB \cite{Liew2011} presents an approach that collects and stores computational data in a database for planning the execution of workflows that occur in memory and out-of-core. 
ParaTrac \cite{Dun2010} is a data-intensive workflow profiler with an interface that allows its use by different SWfMS. ParaTrac uses the Linux {\em taskstats} interface to collect memory usage, CPU and runtime statistics, and the FUSE file system to record the amount of data that is passed between workflow activities. 
In \cite{silvaintegrating}, provenance data is integrated into the TAU code profiling tool, allowing the performance visualization of workflows executed with the Scicumulus SWfMS. This work aims to carry out the monitoring and profiling for the detection of anomalies in the execution of large-scale scientific workflows. 
Visionary \cite{Oliveira2017} is a framework for analysis and visualization of provenance data with the support of ontology and complex networks techniques. However, the work is focused on the analysis of the provenance graph and does not include domain data or predictive analysis. Our approach provides profiling analysis of the execution of workflows from the provenance collected by Swift, in addition to being able to aggregate domain data from the experiment. We show that it is possible to combine provenance and domain data for more detailed analysis. Also, we demonstrate that these analyses can benefit from machine learning techniques for extracting relevant information and predicting computational resource consumption that are not readily detected by queries and statistics from the provenance database.

Focusing on bioinformatics workflows, the WEP \cite{Wep2013} tool enables the modularization of the workflow, allowing the user to execute the entire experiment or just the necessary activities. In addition, WEP also allows access to intermediate files and final results through the web interface. It does not provide detailed computational profiling or parallel and distributed execution features as does BioWorkbench.
ADAM \cite{Adam2013} is a scalable API for genome processing workflows that uses Spark \cite{Zaharia2016}, a framework for managing Big Data. Both tools allow for running experiments in HPC environments, but they implement parallelism at a lower level, within the activities. ADAM requires the API functions to be implemented according to Spark's model of computation, using distributed datasets and applying  actions and transformations, such as map and reduce, to these. BioWorkbench allows for the parallel composition of workflow activities that may be implemented using different models of computation. Therefore, BioWorkbench is more flexible in incorporating existing and legacy bioinformatics tools into its workflows.


\section{Materials and Methods}\label{sec:desimpl}

\subsection{Design and implementation}\label{subsec:desimpl}

\begin{figure*}[!htb]
	\begin{center}
	    \includegraphics[width=0.8\linewidth]{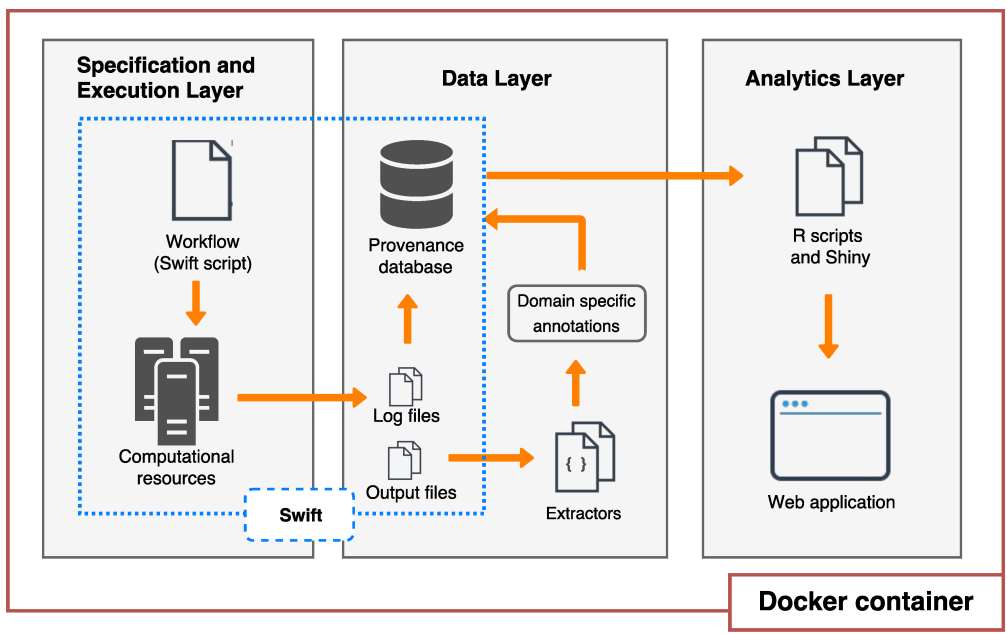}
	\end{center}
    \caption{\label{fig:arq_bio}BioWorkbench layered conceptual architecture.}
\end{figure*}

BioWorkbench 
was developed to support users in the specification process of an experiment, the experiment execution in HPC resources, the management of consumed and produced data, and the analysis of its execution results. 
BioWorkbench uses a SWfMS for definition and execution of bioinformatics workflows, and a web application for provenance data analytics. Figure \ref{fig:arq_bio} shows the layered conceptual architecture of BioWorkbench. 
It should be noted that this architecture is available for demonstration in a Docker \cite{docker} software container\footnote{Available at https://hub.docker.com/r/malumondelli/rasflow/} 
that allows one to reproduce the computational environment described and used in this work. 
In this way, BioWorkbench supports reproducibility at two levels: the computational environment and the data derivation history (through provenance information).
In the following subsections, we detail the features and functionalities of each layer.

\subsection{Specification and Execution Layer}\label{subsec:management}

This layer uses Swift \cite{Wilde2011}, a SWfMS that allows users to specify workflows through a high-level scripting language and execute them in different HPC environments.
The scripting language follows functional programming principles that allow, for example, all operations to have a well-defined set of inputs and outputs and all the variables to be write-once.
Through the Swift scripting language, datasets are declared as variables of primitive types, such as floats, integers, and strings; mapped types, which associate a variable with persistent files; and collections, which include arrays and structures. 
The activities that comprise a workflow are represented as \textit{app functions}, responsible for specifying how applications outside Swift should be executed, their inputs and outputs. \textit{Compound functions} are used to form more complex data flows and can include calls to \textit{app functions}, loops, and conditionals.
All expressions in a Swift script whose data dependencies are met are evaluated in parallel. Through the \textit{foreach} loop instruction, it is possible to process all elements of an array in parallel. 
The independence of locality is another essential feature of Swift and allows the same workflow to run on different computing resources without the need to change its specification. 
Therefore, Swift supports the management and parallel execution of these workflows in HPC environments. 
This layer is then responsible for managing, through Swift, the execution of the workflow in the computational resources to which the user has access and intends to use. 

\subsection{Data Layer}\label{subsec:data}

The data layer handles the provenance of workflow executions.
For this, the layer also uses Swift because of its capability of tracking provenance of the workflow execution behavior as well as its activities \cite{Gadelha2011}. 
All information regarding the execution of a workflow in Swift, or its provenance, is recorded in a log file. Each of the activities executed by a workflow also has a log file, called {\em wrapper log}. These files contain information such as the computational aspects of activity executions: the number of read and write operations to the file system, CPU and memory utilization, and other details. 
Also, the log files keep track of the files consumed and produced by each of the activities, analyzing the data derivation process of the experiment possible. 

\begin{figure*}[htb]
	\begin{center}
	    \includegraphics[width=0.6\linewidth]{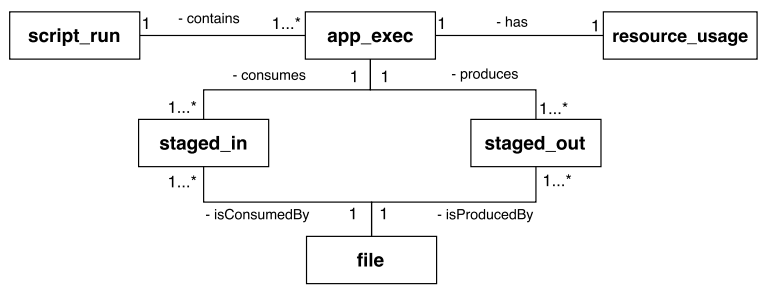}
	\end{center}
    \caption{\label{fig:sqlite}Main entities in the conceptual model of the Swift provenance database.}
\end{figure*}

To provide access to provenance, Swift has a mechanism that processes these log files, filters the entries that contain provenance data, and exports this information to a relational SQL database (SQLite) \cite{Mondelli2016a}.
The main entities of the Swift provenance model are presented in Figure \ref{fig:sqlite}. 
The \textit{script\_run} entity contains the description of the execution of a workflow; the \textit{app\_exec} entity describes aspects of the activities associated with a workflow, and \textit{resource\_usage} has the computational information of the executions of those activities. The \textit{file} entity has the record of all the files consumed and generated by the execution of a workflow; \textit{staged\_in} and \textit{staged\_out} entities, in turn, record the process of deriving data, relating the files to the executions of the activities.

Workflow executions may also produce domain provenance data related to the application area of the experiment. 
Domain data comprises the results obtained by the execution of computational activities of the workflow or parameters used by them. Usually, these results are stored in files. 
In bioinformatics, for example, the domain data from a sequence alignment experiment may consist of rates or scores, indicating to the scientist the accuracy of the alignment. Thus, domain data is data essential for bioinformatics analyses. 
In order to store the domain data along with the provenance Swift collects by default, this layer can use a set of extractors developed for collecting domain data, or annotations, from some bioinformatics applications. 
Extractors consist of scripts that collect domain data from the files produced by the activities, and that can support the analysis process by the user.
These annotations can be exported to the provenance database and associated with the respective workflow or particular activity execution in the form of key-value pairs.
The provenance database contains a set of annotation entities, which relate to the model presented, giving more flexibility to store information that is not explicitly defined in the schema.
More entities can be added to the database allowing for better management of the results of the workflow execution, depending on the type of annotation to be stored. In the section \ref{sec:evaluation}, we describe some of the entities that have been added to the Swift provenance model in order to provide better domain data management of a case study. Once the provenance data is stored in the database, they can be accessed by the {\em Analytics Layer}, described in the next subsection, in order to facilitate access and visualization of the results.

\subsection{Analytics Layer}\label{subsec:vis}

\begin{figure*}[!htb]
	\begin{center}
	    \includegraphics[width=1.0\linewidth]{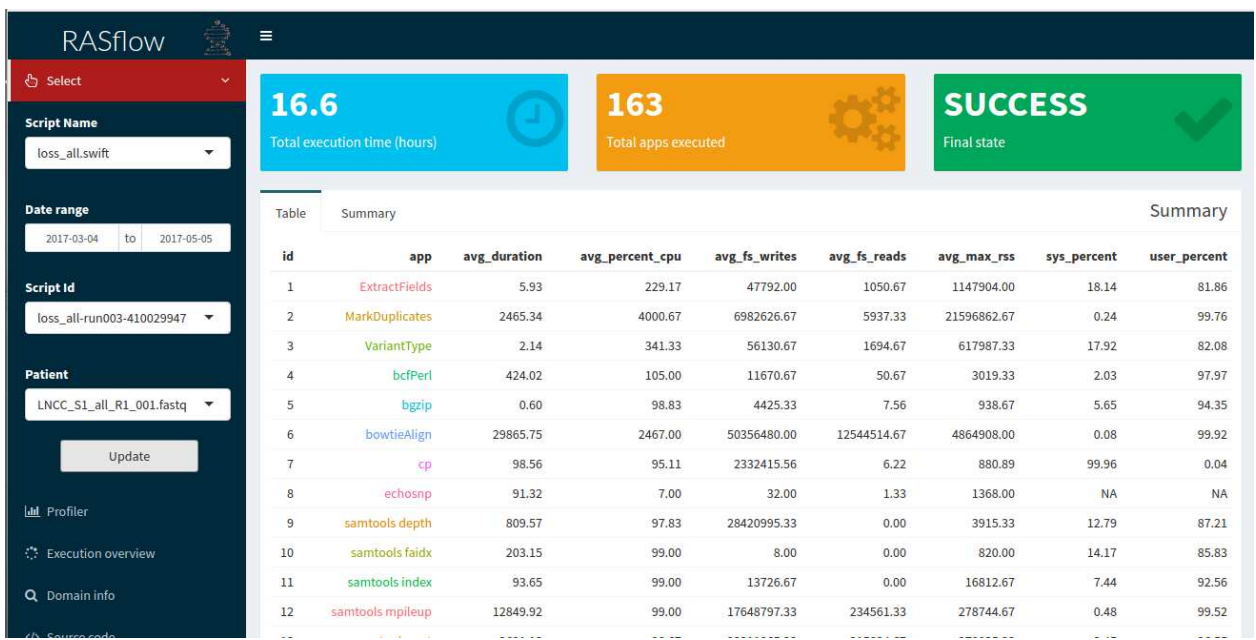}
	\end{center}
    \caption{\label{fig:interface1}BioWorkbench web interface displaying information about a RASopathy analysis workflow (RASflow) execution.}
\end{figure*}

This layer abstracts a set of database queries from the data layer, facilitating the access to the provenance of the workflow. 
The main goal is to provide the user with a more intuitive way to understand the computational profile of the experiment and to analyze the domain results. 
Without the set of query abstractions that we propose in this layer, the user would have to manipulate both the provenance database and the output files of the experiment.
As part of the query abstraction process, we use R scripts that connect to the provenance database, perform the queries and extract useful statistics for analysis. 
The results of the queries retrieve data that are presented to the user through graphs and tables in a web interface. 
This interface was developed using the Shiny library\footnote{https://shiny.rstudio.com/}, which allows an analysis implemented in R to be accessed via web applications interactively. 
The interface provides a menu so that researchers can carry out their analyses. This menu, shown in Figure \ref{fig:interface1}, allows them to select a workflow execution from the list of executions available for analysis. 
The charts and tables are updated according to the chosen options. 
Computational analyses present graphs such as the execution time of the activities, the percentage of CPU used by each of them, and the parallelism level of the workflow. Information about the total time of execution of a workflow, the number of executed activities, and whether the workflow ended successfully is also displayed. 
Domain analyses provide information about the scientific domain of the workflow. 
In the RASflow case study, presented below, this information includes data on the genetic mutations found in patients.

\subsubsection{Machine Learning Techniques in Support of Provenance Analysis}\label{subsubsec:ml}

Provenance data can be a useful resource for a wide variety of machine learning algorithms for data mining. 
The application of machine learning algorithms consists in statistical analyses and automatic generation of models.
These models can produce: (i) predictions, allowing scientists to estimate results or behavior based on some previous information; and (ii) learning, related to the discovery of implicit relations that are not always detected by human observations or simple queries. 
From a data set of interest, the models take into account a set of attributes that must be passed as input to produce an output, which in this case can be a prediction.
If the output receives values in a continuous numeric range, the models comprise a set of mathematical and logical operators, resulting in a regression problem. 
On the other hand, if these outputs can only assume a previously defined set of discrete values, named classes, then the models map the input attributes in classes, resulting in a classification problem. 
In bioinformatics workflows, for example, we can have as input to a machine learning algorithm a set of attributes such as genome sequence size, statistics on memory usage and CPU to analyze them. The output of this algorithm may be the prediction of the execution time. In comparative genomics, more specifically, behaviors or relations such as the required computational time, memory or space for comparing an arbitrary number of genomes can also be predicted.

In this work, as part of the analysis process, we combine provenance data and machine learning by using the free software Weka \cite{hall2009weka}.
Weka provides a broad set of machine learning and statistical analysis techniques for data mining tasks. 
These techniques include rule- and tree-based algorithms, such as \textit{J48, BFTree, OneR}, in which the values of all input and output attributes are previously known. 
To evaluate the performance of a generated model we used metrics related to the differences between the previously known values (which we can call ``real values'') and the predicted values of the output attribute. 
Regarding the generalization capacity of the model, whenever possible the algorithms were evaluated using the methodology known as ``cross-validation,'' with 10 folds \cite{arlot2010survey}.

Therefore, we hope to indicate some relevant opportunities that arise from the association between provenance data and machine learning techniques, both in the specific case of SwiftGECKO, presented below, and in the scientific workflows field. 
Also, these experiments can contribute to the understanding of workflow behavior, being useful, for instance, in guiding optimization efforts and parameter configurations. 
It is worth mentioning that the techniques used in this work were applied using the original algorithm attributes proposed in the software Weka, except for the minimum number of tuples classified at a time by the models. 
This attribute provides a minimum amount of tuples that need to be simultaneously classified by a model for each possible input parameters configuration, thus influencing in the model complexity. The values of this parameter were changed to increase the legibility of the obtained results. 
A more detailed description of the algorithms included in Weka, together with some statistical information that we did not include in this work for scope reasons, can be found in papers such as \cite{arora2012comparative} and \cite{sharma2013weka}.

\section{Results and Discussion}\label{sec:evaluation}

We present three case studies to evaluate BioWorkbench: SwiftPhylo, a phylogenetic tree assembly workflow; SwiftGECKO, a comparative genomics workflow; and RASflow, a RASopathy analysis workflow. 
These workflows are analyzed from both computational and scientific domain points of view using queries to a provenance and annotation database. Some of these queries are available as a pre-built feature of the BioWorkbench web interface. 
The values and statistics presented in this section were extracted from queries to the Swift provenance database. 
In the case of SwiftGECKO and RASflow, in addition to the computational information already collected and stored by Swift, annotations were also gathered from the scientific domain of the experiment, such as the size of the sequences used as input for the workflow execution. 
This demonstrates the usefulness of Swift both in supporting the parallel execution of the experiment and in the analysis of those runs through queries to the provenance database. The executions of each of the workflows were performed on a shared memory computer, with a single node with 160 cores and 2~TB of RAM. 
It is worth mentioning that the computational resource is shared with other users and was not dedicated to these executions. This can be considered as one of the factors influencing the performance gains.
We highlight that the workflows were executed directly in the aforementioned computational resource, not taking into account the Docker structure presented in the Figure \ref{fig:arq_bio}. The Docker container was built for reuse purposes only, to encapsulate the components that compose the framework.

\subsection{SwiftPhylo: phylogenetic tree assembly}

SwiftPhylo\footnote{Available at https://github.com/mmondelli/swift-phylo} is based on the approach proposed in \cite{Ocana2011}. Its goal is to build phylogenetic trees from a set of multi-FASTA files with biological sequences from various organisms. 
The construction of phylogenetic trees allows understanding the evolutionary relationship between organisms, determining the ancestral relationships between known species. Results obtained through phylogenetic experiments contribute, for example, to the development of new drugs \cite{anderson2003process}. 
SwiftPhylo is composed of six activities, shown in Figure \ref{fig:diagram_phylo} and described as follows:

\begin{figure*}[htb]
	\begin{center}
	    \includegraphics[width=0.6\linewidth]{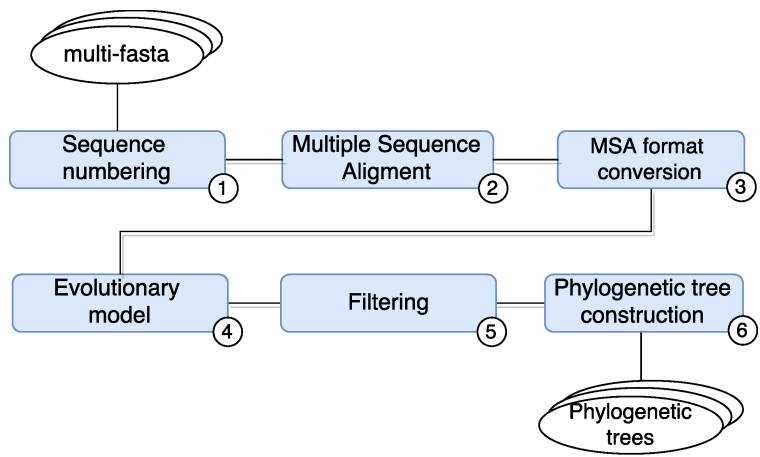}
	\end{center}
    \caption{\label{fig:diagram_phylo}SwiftPhylo workflow modeling.}
\end{figure*}

\begin{enumerate}
\item Sequence numbering: the activity receives multi-FASTA format files as input and uses a Perl script so that each sequence contained in the file receives is numbered.
\item Multiple Sequence Alignment (MSA): the activity receives the result of Activity 1 and produces a MSA as output through the MAFFT \cite{katoh2002mafft} application.
\item Alignment format conversion: this activity converts the format of the file generated by the previous activity to the PHYLIP format using the ReadSeq \cite{Gilbert2003} application.
\item Search for the evolutionary model: this activity tests the output files of Activity 3 to find the best evolutionary model, through the ModelGenerator \cite{keane2006assessment} application.
\item Filtering: this activity uses a Python script to filter the output file from Activity 4.
\item Construction of phylogenetic trees: this activity receives as input the resulting files of Activities 3 and 5, and uses the RAxML \cite{stamatakis2006raxml} application to construct phylogenetic trees.
\end{enumerate}

SwiftPhylo works with a large amount of data and can be run with different parameters. This means that, in practice, managing its execution without the help of a SWfMS becomes a cumbersome task for the scientist. 
Taking into account that the workflow modeling specifies independent processing for each input file, with the implementation of SwiftPhylo we have a workflow that allows us to explore the characteristics of Swift's parallelism.

In the SwiftPhylo implementation process, the computational applications that compose the workflow were mapped to the Swift data flow model. In this way, the activities are represented in the Swift script as \textit{app functions}. The app functions determine how the execution of a computational activity external to Swift is performed, including its arguments, parameters, inputs, and outputs. 
Once this has been done, activity chaining was defined by indicating that the output of one app function is the input of another. Parallelism has been implemented in a way that the activity flow can be executed independently for each input, as shown in the code in Listing \ref{lst:swift_1} through the \textit{foreach} statement.

\begin{figure}[h]
\begin{lstlisting}[label={lst:swift_1}, style=mystyle, frame=single, language=SwiftK, caption={SwiftPhylo specification sample.}]
app (file o) mafft (file i) {
	mafft filename(i) stdout=filename(o);
}

foreach f, i in fastaFile {
	mafftFile[i] = mafft(fastaFile[i]);
}
\end{lstlisting}
\end{figure}

\begin{figure*}[htb]
	\begin{center}
	    \includegraphics[width=0.6\linewidth]{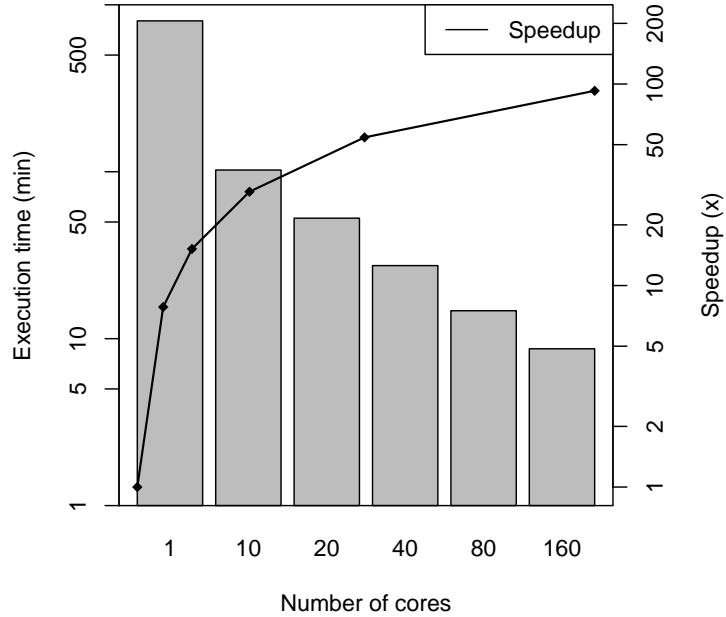}
	\end{center}
    \caption{\label{fig:time_phylo}SwiftPhylo execution time and speedup.}
\end{figure*}

SwiftPhylo was executed using a set of 200 multi-FASTA files, with the size of each file ranging from 2~KB to 6~KB, resulting in the execution of 1200 activities. The total execution time averages and speedup of the workflow are shown in Figure \ref{fig:time_phylo}. The results indicate a reduction of the execution time of the SwiftPhylo from $\sim$13.35 hours, when executed sequentially, to $\sim$8 minutes, when executed in parallel using 160 cores. 
This represents a decrease of 98.9\% in the workflow execution time. Also, the speedup metric was used as a way to evaluate the performance gain between parallel and sequential executions. In this case, SwiftPhylo executed in parallel was $\sim$92 times faster than its sequential execution.

\begin{figure*}[!htb]
	\begin{center}
	    \includegraphics[width=0.8\linewidth]{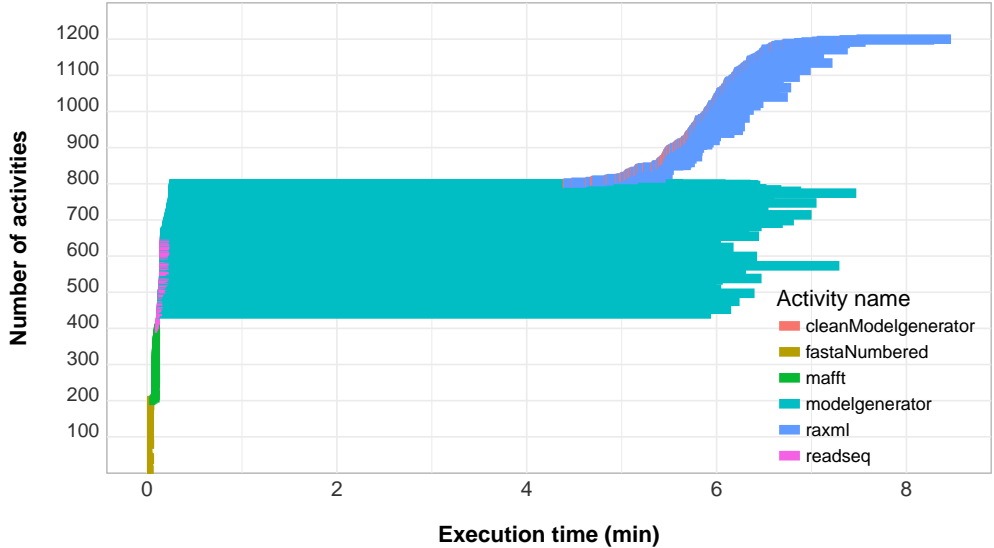}
	\end{center}
    \caption{\label{fig:gantt_phylo}SwiftPhylo workflow Gantt chart expressing its parallelism level.}
\end{figure*}

Another important aspect related to the execution of the workflow concerns its level of parallelism, presented through a Gantt graph in Figure \ref{fig:gantt_phylo}. This type of analysis shows the order of execution and the duration of each of the activities that comprise the \textit{workflow}. In the chart, activities are represented by different colors and each activity is displayed as a horizontal bar that indicates when the activity started and when it ended. 
The bars are stacked on the vertical axis in order of execution. Thus, by plotting a vertical line at some time $t$ of the workflow execution, we find the parallelism at time $t$ as the number of activities intercepted by the line.
We can observe that the {\em modelgenerator} activity is the one that demands the most execution time and can be considered a candidate for identify parallelism strategies that reduce its duration.

\subsection{SwiftGECKO: comparative genomics}

\begin{figure*}[htb]
	\begin{center}
	    \includegraphics[width=0.8\linewidth]{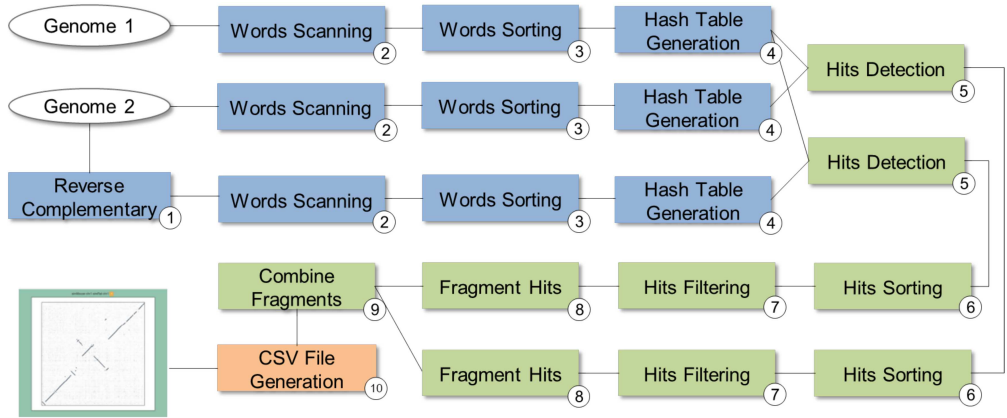}
	\end{center}
    \caption{\label{fig:diagram_gecko}SwiftGECKO workflow modeling.}
\end{figure*}

SwiftGECKO\footnote{Available at https://github.com/mmondelli/swift-gecko} is an implementation of the  comparative genomics workflow proposed in \cite{Torreno2015}. Comparative genomics studies the relationship between genomes of different species, allowing a better understanding of the evolutionary process among them \cite{alfoldi2013comparative}.
SwiftGECKO aims to identify portions of biological sequences of various organisms with a high degree of similarity, or {\em high-scoring segment pairs} (HSP), between them. SwiftGECKO is composed by 10 activities distributed in three modules, which are presented in Figure \ref{fig:diagram_gecko} and described as follows:

\begin{enumerate}
\item Dictionary creation: this corresponds to the creation of dictionaries for each sequence and includes Activities 1 to 4 (blue boxes). For the dictionary creation, the user must indicate the size of the portion of the sequence, or $K$-mers, that will be used for the comparison. The dictionary consists of a table that organizes the $K$-mers and their respective positions, for each of the sequences. 
\item HSP identification: this is composed by Activities 5 to 9 (green boxes) and performs the comparison between the sequences, identifying {\em hits} used to find HSPs. {\em Hits} consist of positions where the $K$-mers of the compared sequences are equal.
\item Post-processing: this is Activity 10, where a conversion of the output format of Activity 9 to CSV is done, allowing analysis of the results. 
The CSV file contains information such as the size of the sequences, the parameters used in the comparison, and the number of hits found.
\end{enumerate}

SwiftGECKO consumes and produces a significant amount of data. A workflow execution with 40 complete bacterial genomes totals 1560 possible comparisons, resulting in the execution of 8080 activities. 
The implementation of SwiftGECKO followed the same steps mentioned in the SwiftPhylo implementation process. Parallelism was divided into two \textit{foreach} stages as shown in Listing \ref{lst:swift_2}. In the first one, the activities that belong to Module 1 are executed; they generate the files that will be consumed by the other activities. The second step, referring to Modules 2 and 3, manages the activities responsible for comparing pairs of all genomes given as input. 

\begin{figure}[h]
\begin{lstlisting}[label={lst:swift_2}, style=mystyle, frame=single, language=SwiftK, caption={SwiftPhylo specification sample.}]
foreach f,i in fasta {  
     wordsUnsort[i] = words(fasta[i]);
}
foreach f,i in fasta {   
     foreach g,j in fasta {     
       hits[i][j] = hits(d2hW[i], d2hW[j], wl);
     }
}
\end{lstlisting}
\end{figure}

SwiftGECKO was executed with a set of 40 files containing bacterial genome sequences, ranging in size from 1~KB to 8~MB. 
The execution time decreased from $\sim$2.1 hours to $\sim$6 minutes, representing a reduction of $\sim$96.6\%. 
Also, the workflow was $\sim$30 times faster when executed in parallel, using 160 processors. In a more in-depth analysis, we illustrate in Table \ref{tab:data_gecko} that the {\em hits} activity is also I/O-intensive. This is a factor that limits the scalability of the workflow.

\begin{table}[!htb]
\centering
\caption{Average duration (s) and the amount of data read and written by SwiftGECKO.}
\label{tab:data_gecko}
\begin{tabular}{|l|r|r|r|}
\hline
\multicolumn{1}{|c|}{\textbf{Activity}} & \multicolumn{1}{c|}{\textbf{Duration (s)}} & \multicolumn{1}{c|}{\textbf{GB read}} & \multicolumn{1}{c|}{\textbf{GB written}} \\ \hline
reverseComplement                        & 11.8                                        & 0.15                                 & 0.14
\\ \hline
words                                    & 55.9                                        & 0.29                                 & 7.05
\\ \hline
sortWords                                & 76.6                                        & 7.05                                & 7.05            
\\ \hline
w2hd                                     & 425.1                                       & 7.05                                & 11.67
\\ \hline
combineFrags                             & 425.8                                       & 0.32                                 & 0.16
\\ \hline
csvGenerator                             & 697.8                                       & 0.006                                  & 0.004                                      \\ \hline
filterHits                               & 2402                                        & 111.36                              & 83.09
\\ \hline
FragHits                                 & 3793.3                                      & 94.7                               & 0.32
\\ \hline
sortHits                                 & 4457.3                                      & 111.36                              & 111.36
\\ \hline
hits                                     & 60058                                       & 455.24                              & 111.36
\\ \hline
\end{tabular}
\end{table}

For a more detailed analysis of the SwiftGECKO provenance data, we present predictive models automatically generated by machine learning methods.
These models aim at understanding how some variables influence the total execution time of the workflow. 
Besides the CPU time, which was selected as the attribute to be predicted, and the total\_fasta\_size, we included the following attributes as input parameters for the models: \textit{length}, \textit{word\_length}, \textit{similarity}, \textit{total\_genomes},  \textit{total\_reads}, and \textit{total\_written}.

The machine learning methods were applied using Weka \cite{hall2009weka}, a software that contains a set of different algorithms for data classification and prediction.
In this case, the regression algorithms obtained especially promising results. 
The simple linear regression algorithm achieved a correlation coefficient of 0.9855 and is represented in the Equation \ref{eq:linear_regression_query4},
while the multilayer neural network algorithm 
obtained a correlation coefficient equal to 0.9941. 
We have not represented this second model due to its high complexity. Commonly, the high complexity of the models obtained by neural network-based algorithms is responsible for the use of the expression ``black box algorithm'', to refer to them.
These results demonstrate that it is quite possible to predict the execution time of the workflow as a whole, on the chosen set of input parameters.

\begin{equation}
\begin{array}{llll}
	CPU\, time=& -931.4025  & \times & \textit{word\_length = 12 $\vee$ word\_length = 10 }\, + \\
			   & -2704.5534 & \times & \textit{word\_length = 10 }\, + \\
    		   & 0       	& \times & \textit{total\_written }\, + \\
    		   & -1006.3049	& \times & \textit{total\_genomes = 20 $\vee$ total\_genomes = 30 $\vee$ total\_genomes = 40 }\, + \\
    		   & -547.6854  & \times & \textit{total\_genomes = 30 $\vee$ total\_genomes = 40 }\, + \\
    		   & 410.1449   & \times & \textit{total\_genomes = 40 }\, + \\
    		   & 876.1037   & & 
\end{array}
\label{eq:linear_regression_query4}
\end{equation}

We divided the domain of possible values for \textit{CPU time} in 5 ranges (which are the classes) $\{ \textit{A, B, C, D, E} \}$ of equal size, in which $A$ includes the lowest values and $E$ includes the highest values. This makes it possible to apply classification algorithms using the same data as the previous analysis.
Having a discrete domain, the information gain ratio values for each attribute are listed in the Table \ref{tab:info_gain_query4}. 
Cost matrices were used to avoid the overvaluation of class A, which holds 88.46\% of all tuples. 
In this case, the tree-based techniques ``J48'' and ``BF Tree'' obtained the best results, correctly classifying  96.79\% and 96.15\%, respectively, of all tuples.

\begin{table}[!h]
\centering
\begin{tabular}{|l|l|}
\hline
\multicolumn{1}{|c|}{\textbf{Information gain ratio}} & \multicolumn{1}{|c|}{\textbf{Attribute}}\\ \hline
1 & total\_read \\ \hline
0.1842 & word\_length \\ \hline
0.1611 & total\_fasta\_size \\ \hline
0.1208 & total\_genomes \\ \hline
0.0421 & length \\ \hline
0.0271 & similarity \\ \hline
\end{tabular}
\caption{Information gain ratio values of the attributes used for the analysis.}
\label{tab:info_gain_query4}
\end{table}

\begin{small}
\begin{algorithm}[h!]
\caption{J48 query4 without constraints}
\label{a:j48-query4-completa}
\If{total\_read $\leq$ 1020229446482}{
	\Return Class A\;
}\Else{
	\If{total\_fasta\_size $\leq$ 113556442}{
    	\If{length = 80}{
        	\If{similarity = 65}{
            	\Return Class C\;
            }\Else{
            	\Return Class B\;
            }
        }\Else{
        	\Return Class C\;
        }
    }\Else{
    	\Return Class E\;
    }
}
\end{algorithm}
\end{small}

\begin{small}
\begin{algorithm}[h!]
\caption{BFTree query4 without constraints}
\label{a:bfTree-query4-completa}
\If{total\_read $<$ 1.768799155948E12}{
	\Return Class A\;
}\Else{
	\If{total\_fasta\_size $<$ 1.314087315E8}{
    	\If{length=100}{
        	\Return Class C\;
        }\Else{
        	\If{similarity $\neq$ 40}{
            	\Return Class C\;
            }\Else{
            	\Return Class B\;
            }
        }
    }\Else{
    	\Return Class E\;
    }
}
\end{algorithm}
\end{small}

The ``OneR'' algorithm obtained the rule given by the Equation \ref{eq:oneR_query4}, correctly classifying 96.79\% of all tuples.
\begin{equation}
\centering
\begin{array}{lll}
	total\_read: & & \\
    	< 1.768799155948E12  & : & Class\, A \\
        < 2.5173879304975E12 & : & Class\, B \\
        < 3.4581148731605E12 & : & Class\, C \\
        \geq 3.4581148731605E12 & : & Class\, D \\
\end{array}
\label{eq:oneR_query4}
\end{equation}

According to the Table \ref{tab:info_gain_query4}, 
the \textit{CPU time} is most directly influenced by the amount of data read by the workflow. 
Together with the third higher information gain ratio value assigned to the attribute \textit{total\_fasta\_size}, we can suggest that the magnitude of the information to be read is the most influential aspect concerning the execution time.
Despite the I/O routines, the attribute \textit{word\_length} receives the second higher information gain ratio value, highlighting its relevance.

In fact, in the absence of data about the number of hits, the attribute \textit{word\_length} assumes the most prominent position to predict the attribute \textit{CPU time}. 
For all this, the value of \textit{word\_length} is also a decisive attribute to predict the execution time of the workflow as a whole, due to its importance in the behavior of the most costly component of the workflow.

Reinforcing the relevance of the I/O routines, both Algorithms \ref{a:j48-query4-completa} and \ref{a:bfTree-query4-completa} employ the attributes \textit{total\_read} and \textit{total\_fasta\_size} in their main conditionals. 
As in the Equation \ref{eq:oneR_query4}, the greater the \textit{total\_read} value, the greater the computational time demanded. 
Also, the attributes \textit{length} and \textit{similarity} are used to classify a minor number of tuples. However, they can indicate the influence of the cost associated with the amount of fragments found (influenced by the attributes \textit{length} and \textit{similarity}).

Beyond all this, these experiments demonstrate that is possible to generate simple models able to efficiently predict the computational time demanded by the workflow execution, both in terms of continuous or discrete times.
The techniques used to build these models also allows us to infer predictions related to other attributes, beyond the execution time. For example, we can focus on the amount of written data by the workflow or the specific domain data stored in the provenance database.
Also, there are a wide variety of machine learning methods that produce symbolic solutions and allow, in addition to the predictions, knowledge extraction about different aspects. 
These aspects include, for example, the structure or the importance degree of relations among the various variables or between variables and constants.
Therefore, machine learning methods constitute an essential toolkit to be explored in provenance data analyses concerning SwiftGECKO and other, to provide predictive models and to reveal implicit knowledge.

\subsection{RASflow: RASopathy analysis}

Genetic diseases, such as RASopathies, occur due to changes in the nucleotide sequence of a particular part of the DNA. These changes modify the structure of a protein, which may cause anatomical and physiological anomalies \cite{klug1997}. RASopathies comprise a set of syndromes characterized by heterogeneity of clinical signs and symptoms that limit the prognosis, still in the childhood, of predispositions to certain tumors or neurocognitive delays. RASopathies are characterized by mutations in genes that encode proteins of the RAS/MAPK cell signaling pathway, which is part of the communication system responsible for the coordination of cellular activities and functions. This type of mutation was found in 20-30\% of cancer cases in humans, setting the RASopathies in the group of syndromes that are predisposed to cancer \cite{Lapunzina2014}. 

A consortium between Bioinformatics Laboratory of the National Laboratory for Scientific Computing (LABINFO/LNCC) and the Center for Medical Genetics of the Fernandes Figueiras Institute at Fiocruz (IFF/Fiocruz) established an experiment aimed at the molecular study of RASopathies, from the sequencing of the exome of a set of patients. The consortium seeks to investigate genetic aspects of RASopathies by using DNA sequencing technologies, providing support for the treatment of the disease. 
For this, a study that counts on the processing of the sequences of patients diagnosed with RASopathies was established. This study comprises the use of a set of bioinformatics applications and, due to the significant amount of data to be processed, demands high-performance computing.

\begin{figure*}[!htb]
	\begin{center}
	    \includegraphics[width=0.6\linewidth]{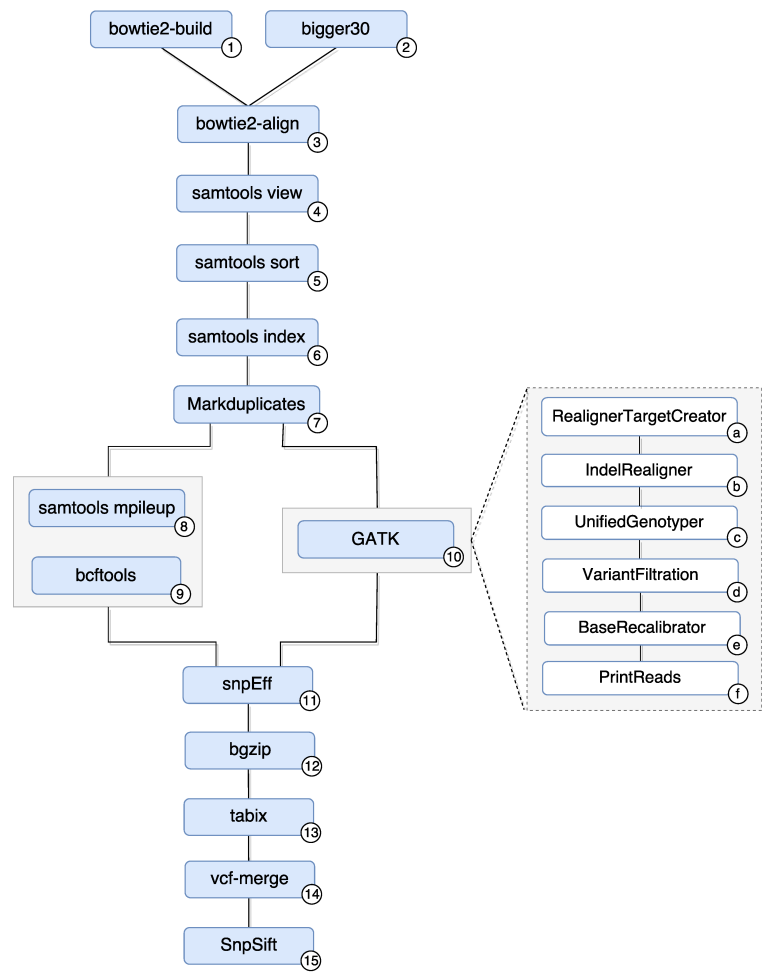}
	\end{center}
    \caption{\label{fig:modeling}RASopathy analysis workflow modeling.}
\end{figure*}

A scientific workflow, called RASflow, was implemented to support the large-scale management of a bioinformatic experiment to analyze diseases associated with RASopathies and allows the identification of mutations in the exome of patients. Once the exomes of the collected patient samples are sequenced, the results are stored in a text file in the FASTQ format, which records the {\em reads} and their quality scores. The {\em reads} consist of sequence fragments generated by the sequencer. 
By using the results obtained through RASflow, a researcher can analyze and identify whether or not there is pathogenicity in the mutations present in the exome.
RASflow was implemented in Swift, and the workflow model is shown in Figure \ref{fig:modeling}.

Activity 1 receives a reference human genome and indexes it. The results are stored in files in the BT2 format, which records the indexes according to the sequence size. This activity requires large computational time, and its results are used in the analysis of each patient. Therefore, the workflow checks for the existence of BT2 files in the filesystem where it is executed. The activity is only performed when the files are not found.

Activity 2 receives the exome sequence of the patient in the FASTQ format as input, filters the reads with a certain quality, and maintains the result in the same format. The need to perform Activity 2 depends on the type of sequencer used to sequence the exome of the patient. If the Illumina sequence was used, for example, it is not necessary to perform this activity. If the sequencing comes from IonTorrent, the activity is performed in parallel.

Activity 3 receives the result of Activities 1 and 2 to perform the alignment of the exome and the reference genome sequence, resulting in a SAM format file and a log file with alignment statistics. The SAM file, the result of the alignment, is usually large and therefore difficult to analyze and manipulate. Activities 4, 5 and 6 are used to compress, sort and index the SAM file, respectively, generating the BAM and BAI binary files that store the aligned sequences. The resulting files are used by Activity 7, which filters duplicate reads and produces a file in BAM format as well.

From this point on, mutations are identified by comparing the exome sequence of the patient and the reference genome. However, the workflow presents some variability, and the comparison can be made through two approaches: (i)
using {\em samtools} \cite{samtools2011} or (ii) the toolkit provided by \textit{GATK}  \cite{gatk2010}. A more detailed analysis of the differences between the two approaches that take into account biological aspects falls outside the scope of this work. However, it can be said that the GATK toolkit comprises more sophisticated techniques and the results obtained tend to be more accurate.

In the workflow variation that uses \textit{samtools}, the BAM file produced by Activity 7 and the reference genome are consumed by Activity 8, which performs the comparison. The result is used by Activity 9, which converts the file from a binary format to the VCF format, which is responsible for storing the genetic variants found. The workflow variation using the \textit{GATK} approach performs the set of Activities 10a-10f, indicated in the figure, which also produces a VCF file. The VCF file resulting from the execution of one of the variations is consumed by Activity 11 to predict the effects of the mutations found and to determine the pathogenicity, resulting in a set of VCFs. In this work, we also included Activities 12 and 13 to compress and index the files resulting from Activity 11, respectively. The results are consumed by Activity 14, responsible for combining this set of files. Finally, Activity 15 applies a set of filtering activities based, for example, on the quality and quantity of occurrences of a given variant, generating a single final VCF file.

\begin{figure*}[!htb]
	\begin{center}
	    \includegraphics[width=0.5\linewidth]{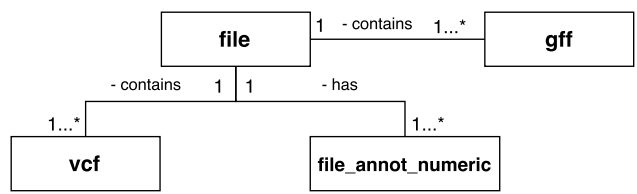}
	\end{center}
    \caption{\label{fig:domainrasflow}Database entities for storing scientific domain annotations in RASflow.}
\end{figure*}

A set of scientific domain annotation extractors was developed for gathering and adding to the provenance database information from the following files: (i) the GFF file, which describes characteristics of the DNA, RNA and protein sequences of the genome used as reference; (ii) the log file resulting from the execution of Activity 3; and (iii) the file in VCF format, which contains the variations found in the exome. 
The conceptual schema of the Swift database has been augmented to enable storage of this information. The following tables, as shown in Figure \ref{fig:domainrasflow}, were added: \textit{gff} and \textit{vcf}, for storing the contents of the GFF and VCF files respectively; and the \textit{file\_annot\_numeric} table, which records the contents of the log file in key-value format. 

Next, we propose a set of queries to the RASflow provenance database to assist the scientist in the analysis of the results of the workflow. 
The queries were defined with one of the researchers responsible for the analysis of RASopathies.
They are available for viewing in the BioWorkbench web interface corresponding to the RASflow experiment. That is, the scientist does not need to redefine them to obtain the results.

\textit{Query 1}: ``Retrieve the alignment rate of each patient's exome sequences relative to the reference genome''. 
The alignment process allows identifying the similarity between the patient sequence and the sequence used as a reference. The result is obtained by executing Activity 2 of the workflow, which produces a log file. This file records some statistics and among them the rate of alignment between the two sequences. This type of information is useful for the scientist, who is responsible for deciding what rate is sufficient for proceeding with the analysis. The alignment rate is stored in the \textit{file\_annot\_numeric} table of the provenance database and can be retrieved through the query shown in Listing \ref{lst:query_1}. Table \ref{tab:query_1} displays the result of the query, with the alignment rates for each of the patients used in the analysis of RASopathies of a given workflow execution.

\begin{figure}[h]
\begin{lstlisting}[label={lst:query_1}, style=mystyle, frame=single, language=SQL, caption={SQL query to retrieve the alignment rate of each patient's exome sequences.}]
SELECT         file_id, value 
FROM           file_annot_numeric
NATURAL JOIN   staged_out
WHERE key LIKE 'overall alignment rate' AND   
               app_exec_id LIKE '%loss_all-run003-410029947%';
\end{lstlisting}
\end{figure}

\begin{table}[!htb]
\centering
\begin{tabular}{|l|r|}
\hline
\textbf{Patient} & \textbf{Alignment rate} \\ \hline
P1.log            & 93.95\%                      \\ \hline
P2.log            & 94.52\%                      \\ \hline
P3.log            & 94.41\%                      \\ \hline
P4.log            & 94.48\%                      \\ \hline
P5.log            & 94.62\%                      \\ \hline
P6.log            & 94.58\%                      \\ \hline
\end{tabular}
\caption{Alignment rate resulting from patients analysis in RASflow.}
\label{tab:query_1}
\end{table}

\textit{Query 2}: ``Retrieve the biotype and transcript gene name from the final VCF file''. 
This type of result, without BioWorkbench support, was obtained through a manual scan of the GFF file, used as input in the execution of the workflow, and if the VCF file generated as the final result. In RASflow, these two files are imported into the \textit{gff} and \textit{vcf} tables of the provenance database. To retrieve this information, one can use the query presented in Listing \ref{lst:query_2}, which performs the join of the \textit{vcf} and \textit{gff} tables for a given workflow execution. The result is displayed in Table \ref{tab:query_2}. Through the BioWorkbench interface for RASflow, the result is also presented as a table, but iteratively, allowing the scientist to filter the columns or search for names according to their needs.


\begin{figure}[h]
\begin{lstlisting}[label={lst:query_2}, style=mystyle, language=SQL, frame=single, caption={SQL query to retrieve the biotype and transcript gene name from the final VCF file of a patient.}]
SELECT       v.*, g.Name AS nome, 
             g.biotype AS biotipo
FROM         vcf v
NATURAL JOIN file f
NATURAL JOIN staged_out o
NATURAL JOIN app_exec a
LEFT JOIN    gff g ON v.trid = g.ID
WHERE        a.script_run_id LIKE 'loss-run006-3737171381';
\end{lstlisting}
\end{figure}

\begin{table*}[!htb]
\centering
\begin{tabular}{|l|l|l|l|l|}
\hline
\textbf{Patient}                          & \textbf{Gene}   & \textbf{Transcript} & \textbf{Name} & \textbf{Biotype} \\ \hline
P1.log & ENSE00001768193 & ENST00000341065     & SAMD11-001    & protein\_coding  \\ \hline
P1.log & ENSE00003734555 & ENST00000617307     & SAMD11-203    & protein\_coding  \\ \hline
P1.log & ENSE00003734555 & ENST00000618779     & SAMD11-206    & protein\_coding  \\ \hline
P1.log & ENSE00001864899 & ENST00000342066     & SAMD11-010    & protein\_coding  \\ \hline
P1.log & ENSE00003734555 & ENST00000622503     & SAMD11-208    & protein\_coding  \\ \hline
\end{tabular}
\caption{Mutation list with the biotype and the name of the transcribed genes in the final VCF file of a patient.}
\label{tab:query_2}
\end{table*}

To evaluate the performance of the workflow, we used exome sequences of 6 patients as input, ranging in size from 8~GB to 11~GB. 
It is worth mentioning that a preliminary version of this workflow was developed in Python. In this work, we chose to develop it using Swift to take advantage of the parallelism offered by the system.
By considering the variability of the workflow, executions were made for the two possible approaches: one using the \textit{samtools} tool and the other using the \textit{GATK} toolkit. 
In the approach using \textit{samtools}, a reduction of $\sim$77\% of the execution time of the analyses was obtained, representing a gain of $\sim$4 times in the execution with Swift compared to the sequential execution in Python. The approach using \textit{GATK} had a reduction of $\sim$80\% in execution time, for a gain of $\sim$5 times. The parallelism strategy of RASflow explored the simultaneous analysis of patients. Thus, the total time of execution of the workflow is associated with the execution time of the most time-consuming patient analysis. 

Because it has been executed with a small set of patients, the workflow has a large run-time variability. 
However, genetic variant analysis can be applied to other diseases.
For diseases such as cancer, for example, the volume of sequenced genomes is much larger and therefore, there is an opportunity to obtain higher performance gain through the approach used in RASflow.

\section{Conclusion and Future Work}\label{sec:conclusion}

Large-scale bioinformatics experiments involve the execution of a flow of computational applications and demand HPC. The execution management of these experiments, as well as the analysis of results, requires a lot of effort by the scientist. 
In this work, we demonstrate that the use of scientific workflow technologies coupled with provenance data analytics can support this management, allowing for the specification of experiments, parallel execution in HPC environments, and gathering provenance information.

To benefit from the use of scientific workflow technologies and support the whole process of scientific experimentation, we have developed the BioWorkbench framework. It was designed to use the Swift SWfMS for the specification and execution of bioinformatics experiments, and a web application for provenance data analytics.
In this way, through BioWorkbench the user has access to a tool that integrates various features ranging from the high-performance execution of the workflow to profiling, prediction, and domain data analysis.
We can observe that BioWorkbench enables a better scientific data management since the user does not have to directly manipulate the provenance database and the resulting files from the experiment execution. Another important aspect concerns the reproducibility of the experiment, which is facilitated by the provenance and the reproduction of the computational environment through a Docker container. 

We used three case studies that model bioinformatic experiments as workflows: SwiftPhylo, SwiftGECKO, and RASflow. In addition to the performance gains achieved by using Swift in BioWorkbench, we have demonstrated how the provenance allows the identification of bottlenecks and possible optimization opportunities in the execution of workflows. Also, we conclude that users can benefit from the application of machine learning techniques in provenance analysis to, for example, predict and classify workflow execution time. It is noteworthy that, during the development of this work, SwiftGECKO was integrated into the Bioinfo-Portal scientific portal\footnote{Available at http://bioinfo.lncc.br/} \cite{Mondelli2016a}. Through the portal, users can execute the workflow through a web interface in a transparent way, using geographically distributed computational resources managed by the Brazilian National System for High-performance Computing (SINAPAD).

\section*{Acknowledgements}

This work was partially supported by Brazilian funding agencies CNPq, CAPES, and FAPERJ. M. Wilde has an employment and ownership interest in the commercial firm Parallel Works Inc.

\bibliographystyle{abbrv}

\end{document}